\documentstyle[preprint,prl,aps]{revtex}
\begin{document}
\draft
\title{unusual temperature dependent resistivity of a semiconductor 
quantum wire}
\author{Lian Zheng and S. Das Sarma} 
\address{Department of Physics,
University of Maryland, College Park, Maryland 20742-4111}
\date{\today}
\maketitle
\begin{abstract}
We calculate the electronic resistivity
of a GaAs-based semiconductor quantum wire 
in the presence of acoustic phonon scattering.
We find that the usual Drude-Boltzmann transport theory leads to 
a low temperature activated behavior instead of the 
well-known Bloch-Gr\"uneisen power law. Many-body electron-phonon 
renormalization, which is entirely negligible in 
higher dimensional systems,
has a dramatic effect on the low temperature quantum wire transport
properties as it qualitatively modifies the temperature dependence 
of the resistivity from 
the exponentially activated behavior 
to an approximate power law behavior at sufficiently low temperatures.
\end{abstract}
\pacs{72.20.Fr 71.10.Pm 73.40Dx 71.38+i}
\narrowtext
Much recent interest has focused on electronic properties of semiconductor 
quantum wire structures, particularly in the extreme quantum limit 
where the electrons are essentially confined in one dimensional channels.
One-dimensional quantum wire systems possess many unique properties.
One of them is the suggested high mobility \cite{sak1} for
electrons
at low temperatures.
Because of the severe phase-space restriction,  
the only possible low energy excitations of a one-dimensional
electron gas (1DEG) have wavevectors of magnitude $0$ or $2k_F$,
where $k_F$ is the 1D Fermi wavevector.
If one considers the low temperature momentum relaxation of 
a 1DEG, it has to occur through the transfer of momentum $2\hbar k_F$
from the electrons to 
phonons or impurities. 
In the case of electrons coupled to acoustic phonons, 
the situation addressed in this letter,
which have a linear dispersion of the form $\omega_q=cq$, 
the resistive scattering events occur  
with a characteristic energy scale of $\hbar \omega_{2k_F}=2\hbar ck_F$.
One would expect that when $k_BT\ll\omega_{2k_F}$,
where $T$ is the temperature and $k_B$ is the Boltzmann constant,
the resistivity of an electron gas interacting with 
acoustic phonons would follow an exponential behavior 
exp$(-\hbar \omega_{2k_F}/k_BT)$, simply because the
low temperature scattering is entirely mediated by 
phonons of energy $\hbar \omega_{2k_F}$.
Thus, in a 1DEG there is no conventional Bloch-Gr\"{u}neisen  
temperature regime in the 
resistivity---all low temperature acoustic phonon scattering processes
are exponentially activated in a 1DEG.
The usual Drude-Boltzmann transport theory therefore leads to 
activated transport behavior at low temperature even in free carrier 
``metallic'' quantum wires!  This is in striking contrast to 
higher-dimensional systems.
In this letter, we report our theoretical 
study on the resistivity of an interacting 
1DEG in the presence of acoustic phonon scattering using the
memory-function formalism. We show
that the above Drude-Boltzmann conclusion of an exponentially activated 
resistivity due to phase-space restriction 
is in fact incorrect at sufficiently
low temperatures. 
Phonon renormalization,
which is inevitable in the presence of the electron-phonon 
interaction, enhances the resistive scattering rate at low temperatures
and results in a power law 
temperature dependence of the resistivity at sufficiently low temperatures. 
This dramatic effect of phonon renormalization is shown to be unique 
to 1DEG with no corresponding higher dimensional analogy.
Although there exists a substantial 
body of theoretical work \cite{1dr} on the
transport properties of quantum wires, 
the phenomenon discussed in this
letter has not been predicted before to the best of our knowledge.

In a coupled electron-acoustic phonon many-body system, the properties of
the electrons and the phonons are mutually coupled through
electron-phonon interaction.  
In addition to renormalizing the bare phonon frequency, the interaction 
also results in a finite lifetime for the phonons,
{\it i.e.}, a broadening of the phonon spectral function.  
This means that the renormalized frequency of the phonons interacting 
with the electrons is not a sharply defined quantity, but rather a 
resonance of finite width around the bare phonon frequency: 
there is a nonzero amplitude for the renormalized phonons 
to have arbitrary frequencies
for a given wavevector. 
As a consequence, the resistive scattering of the 1DEG by the 
acoustic phonons, which is restricted to have a wavevector $2k_F$
from the phase-space consideration, can occur through phonons
not just with the bare frequency $2ck_F$, but with all possible frequencies, 
in particular, with arbitrarily low frequencies. 
Strictly speaking, the simple exponentially 
activated low temperature behavior
of the resistivity suggested above, which is 
based on the existence of a single 
phonon energy scale of $2c\hbar k_F$, is lost.
The questions that need to be addressed now are what 
quantitative effect the phonon renormalization has
and how significant it is.
Firstly, the phonon renormalization
increases the resistivity 
compared with the bare phonon situation at low temperatures
because the scattering from the
low energy phonons located at the tail of the 
broadened phonon spectrum, which will be
referred to as virtual phonons hereafter,  
is not exponentially suppressed at any non-zero temperature
because low energy renormalized phonons are always available.
Secondly, the enhancement of the resistivity 
becomes very important at sufficiently low temperatures
for one-dimensional electron systems. 
The electron-phonon interaction creates arbitrarily low energy
virtual phonons, but these virtual phonons carry a very tiny fraction
of the total phonon spectral weight. Most of the spectral weight 
is still retained by the bare phonon,
so that the broadening is completely unimportant at high temperatures,
and at any temperature in higher dimensional systems. 
Because phase-space consideration in 1D severely restricts
the wavevectors of the scattering phonons to $2k_F$, 
the contribution to the resistivity from the bare phonons
becomes effectively frozen out at $k_BT\ll\hbar\omega_{2k_F}=2\hbar ck_F$, 
leaving the scattering from the virtual phonons 
as the only remaining contribution to the resistivity
of quantum wires at low temperatures.
Many-body renormalization of the phonon spectral function thus
becomes the dominating factor in determining the low temperature 
resistivity. 
For higher dimensional electron systems, 
the restriction on the wavevector of the phonons in the scattering
events is not present, so the scattering by 
phonons of arbitrarily small wavevectors is allowed. The contribution
from the bare phonons is not frozen out at any finite temperatures
in higher dimensions.
Due to the large spectral weight of the bare phonons, their
contribution to the resistivity therefore always dominates  
in higher dimensions with virtual phonon scattering being
negligibly (orders of magnitude) small. Thus, 
in higher dimensional electron systems the many-body phonon renormalization 
is an entirely negligible effect, 
and the familiar 
Bloch-Gr\"{u}neisen behavior \cite{pri} remains valid
for phonon scattering.

The above discussion can be represented by Fig. \ref{f1},
where the excitation spectrum of a 1DEG at zero temperature is shown
along with the bare phonon dispersion. The most distinct feature 
for a 1DEG is that the low energy part of the phase-space,  
the region below the dotted-line $BCD$ in Fig. \ref{f1}, is excluded
from the particle-hole excitation spectrum. 
The acoustic phonon dispersion line and the particle-hole excitation 
region overlap only in a narrow strip around $q\sim2k_F$, 
where the electron-phonon scattering can take place. 
This is not the case in higher dimensions. The particle-hole 
excitation spectrum for a two- or
three-dimensional electron gas, for example,
contains all the low energy regions between the dotted lines $AB$ and $DE$
in Fig. \ref{f1}.  The phonon dispersion line with $q\leq2k_F$ is 
completely within the particle-hole excitation region, 
so electron-phonon scattering can occur for arbitrary 
wavevectors $q\leq2k_F$.
Due to the large spectral weight carried by the bare phonon, the 
many-body broadening of the phonon spectrum is always negligibly
small in higher dimensions.

In the following, we first briefly describe our calculation and 
then present the results. 
The resistivity of a 1DEG with acoustic phonon scattering 
is calculated by using the
memory function formalism 
\cite{gp,for,find,dc,lz}.
One significant advantage of the memory function formalism 
is that vertex corrections are automatically incorporated in the theory,
and one does not need to work with complicated 
electron-phonon vertex equation \cite{hol}. 
In the memory function formalism \cite{gp,for,find},
the resistivity is given by
\begin{equation}
\rho=-{1\over\hbar n^2e^2}\lim_{\omega\rightarrow0}
{{\rm Im}R^{ret}(\omega)\over\omega},
\label{equ:r1}
\end{equation}
where $n$ is the average density of the 1DEG 
and $e$ is the electron charge.
$R^{ret}(\omega)$ is the retarded force-force
correlation function 
\begin{equation}
R(t)=-{i\over L}< [F(t),F]>,
\label{equ:r2}
\end{equation}
where $L$ is the length of the 1DEG. For phonon scattering the force 
operator $F$ is
\begin{equation}
F=\sum_q iq \varrho(q){M_q\over L^{1/2}}(a_{-q}+a^\dag_q),
\label{equ:ff1}
\end{equation}
where $\varrho(q)$ is the density operator of the 1DEG, 
$a_q(a_q^\dag)$ is the phonon annihilation (creation) operator,
$M_q$ is the electron-acoustic-phonon deformation 
potential coupling matrix element
\cite{1d1p}:
$M_q^2=\hbar D_s^2q^2/(2\varrho_sab\omega_q)$, 
where $D_s$ is the deformation potential, 
$\varrho_s$ is the density of the 
substrate material, $a$ and $b$ are the widths of the 1DEG wire,
which is taken to be of rectangular cross-section within the infinite well 
confinement approximation \cite{1d1p}. 
Combining Eq.(\ref{equ:r1})-(\ref{equ:ff1}), and after a
standard procedure of contour integration and analytic continuation,
we obtain the following expression for the 
acoustic phonon scattering resistivity
\begin{equation}
\rho={\hbar\beta\over2\pi^2n^2e^2}\int^\infty_0dqq^2|M_q|^2
\int^\infty_0 d\omega{{\rm Im}\chi^{ret}(q,\omega)
{\rm Im}D^{ret}(q,\omega)\over
{\rm sinh}^2(\hbar\beta\omega/2)},
\label{equ:r3}
\end{equation}
where $\beta=1/(k_BT)$,
$\chi^{ret}(q,\omega)$ is the retarded 
density-density response function of the 
1DEG, and $D^{ret}(q,\omega)$ is retarded phonon propagator.
In the random-phase approximation (RPA), 
$\chi(q,\omega)=\chi_o(q,\omega)/[1-v_q\chi_o(q,\omega)]$,
with the finite temperature polarizability $\chi_o(q,\omega)$
of a free 1DEG obtained
from its zero-temperature counterpart by an
integration over chemical potential \cite{hus}.
For a 1DEG with finite wire widths,
the Coulomb interaction potential is \cite{1d1p}
$v_q=(e^2/\epsilon_s)\int d\eta |I(\eta)|^2H(q,\eta)|
/(q^2+\eta^2)^{1/2}$, where $\epsilon_s$ is the dielectric constant
of the substrate material. The form factor $I(\eta)$ and
$H(q,\eta)$ used in our calculation are taken from
the infinite well confinement model \cite{1d1p}.
The phonon propagator is
\begin{equation}
D(q,\omega)={2\omega_q\over\omega^2-\omega^2_q
-2\omega_q|M_q|^2\chi(q,\omega)}.
\label{equ:d1}
\end{equation}
The last term in the denominator is the phonon self-energy
arising from the electron-phonon interaction.
This self-energy gives rise to the broadening of the phonon 
spectrum, whose effect is the main concern of the present work. 
When the phonon renormalization is ignored, one has
Im$D^{\rm ret}(q,\omega)=$Im$D^{\rm ret}_0(q,\omega)
=\pi[\delta(\omega+\omega_q)-\delta(\omega-\omega_q)]$
with $\omega_q$ being the bare acoustic phonon frequency.
Inserting this into Eq. (\ref{equ:r3}), one obtains 
the Drude-Boltzmann resistivity due to bare phonons
\begin{equation}
\rho_o={\hbar\beta\over2\pi n^2e^2}\int^\infty_0dqq^2|M_q|^2
{(-1){\rm Im}\chi^{ret}(q,\omega_q)\over
{\rm sinh}^2(\hbar\beta\omega_q/2)}.
\label{equ:r4}
\end{equation}
The characteristic of the bare phonon resistivity is 
an approximate exponential temperature dependence 
$\rho_o\propto{\rm exp}(-\hbar\omega_{2k_F}\beta)$,
which comes from the hyperbolic function 
${\rm sinh}^2(\hbar\beta\omega_q/2)$ in the above expression. 
The effect of many-body broadening of the phonon spectrum
can be inferred by comparing the results
of Eq. (\ref{equ:r3}) and
(\ref{equ:r4}).

The numerical results of our calculation 
are presented in Fig. \ref{f2} to \ref{f4},
where we take the parameters which are appropriate for 
semiconductor quantum wires in GaAs materials:
deformation potential $D_s=7.0$\ eV, density $\varrho_s=5.307$\ gcm$^{-3}$,
speed of sound $c=4.73\times10^5$\ cms$^{-1}$, effective mass
$m/m_o=0.069$, dielectric constant $\epsilon_s=12.9$.
In Fig. \ref{f2}, the results for calculated resistivities,
expressed in terms of mobilities $\mu=(ne\rho)^{-1}$, 
are shown as functions of temperature $T$.
The bare phonon result of Eq. (\ref{equ:r4}), 
the dotted-line in Fig. {\ref{f2},
shows an approximate exponential temperature dependence 
as we mentioned above.
In addition to the bare phonons, the broadened phonon spectrum 
also contains the virtual phonons. The bare phonons carry
large spectral weight and have high energy, while the virtual phonons
carry small spectral weight but have arbitrarily low energy.
At high temperatures ($k_BT\geq\hbar\omega_{2k_F}$), the bare 
phonon contribution dominates because of the large spectral weight. 
The total resistivity 
is essentially the same as that of the bare phonons.
As the temperature is lowered, the resistivity 
due to the bare phonons drops exponentially and 
the contribution from the virtual phonons begins to dominate.
The temperature where the 
deviation from the bare phonon result starts to become appreciable is 
$k_BT/\hbar\omega_{2k_F}<0.1$, which is $T<1K$ for a 1DEG of 
density $n=10^6{\rm cm}^{-1}$.
The temperature dependence of the total resistivity, 
including the phonon renormalization effect,
in the low temperature regime
is shown in the insert of Fig. \ref{f2}. 
One can see that an approximate power 
law behavior $\rho\sim T^\nu$ is found, with $\nu\sim 1.15$ indicated by
the slope of the curve in the insert. 
It should be noted that it is only an approximate
power law dependence. The exact temperature dependence is actually
very complicated because of the logarithmic divergence of the 
density-density response function of a 1DEG at zero temperature. 
This low temperature behavior of resistivity can also be inferred 
analytically by 
performing the $T\rightarrow0$ expansion on Eq. (\ref{equ:r3}).
 
The relative importance of the contributions to the resistivity
from the bare phonons and from the virtual phonons is
compared in Fig. \ref{f3} at high 
($k_BT\sim\hbar\omega_{2k_F}$) and low temperatures 
($k_BT\ll\hbar\omega_{2k_F}$),
where $I(\omega)$ is defined by rewriting 
Eq. (\ref{equ:r3}) as \linebreak[4]$\rho=\int_0^\infty\ I(\omega)d\omega$.
It is clearly seen that the contribution 
to the resistivity is dominated by the bare phonons 
with $\omega\sim \omega_{2k_F}$ at high temperatures 
and by the virtual phonons with $\omega\ll \omega_{2k_F}$
at low temperatures. 
Finally in Fig. \ref{f4}, we show the 
calculated resistivities from the  
bare phonons and
from the renormalized phonons as functions of the quantum wire 
carrier density 
in the low temperature regime.
The bare phonon result shows an exponential dependence
on the density because of $\rho_o\sim e^{-2\beta\hbar ck_F}$,
while the renormalized phonon result shows a very weak dependence
on the density on the scale of this figure.
Again, as in the temperature dependence of the resistivity,
the renormalized phonon theory gives the usual
``metallic'' behavior of weak density dependence,
whereas the usual Drude-Boltzmann transport theory,
which includes only scattering by bare phonons, leads to
unphysically strong exponential density dependence.

We conclude by summarizing our rather dramatic finding with respect 
to the low temperature acoustic phonon scattering 
limited resistivity of 1D semiconductor quantum wire structures:
The usual Drude-Boltzmann transport theory,
which considers scattering by bare phonons only, results in 
extremely ``non-metallic'' exponentially strong temperature 
and density dependence of low
temperature resistivity due to severe 1D phase space restrictions, 
whereas a more refined theory, which incorporates 
many-body renormalization of the phonon spectral function,
restores the usual 
Bloch-Gr\"uneisen power law temperature dependence and weak 
density dependence in the low temperature resistivity. 
We believe that we have discovered a very peculiar and rather subtle
purely one dimensional many-body phenomenon 
which should be experimentally
observable. 

This work is supported by the
U.S.-ARO and the U.S.-ONR.

\begin{figure}
\caption{Excitation spectrum of a
one-dimensional electron gas at zero temperature.  
The particle-hole excitations are allowed in the region
surrounded by the dotted line $ABCDE$. 
The acoustic phonon dispersion is shown as the solid line. 
}
\label{f1}
\end{figure}

\begin{figure}
\caption{Resistivities of a quantum wire
from bare phonon scattering (dotted-line) and from
the renormalized phonon scattering (solid line) 
as functions of temperature. 
The inset shows the power-law behavior of the resistivity 
from the renormalized phonon scattering in the low temperature regime.
The density of the electron gas is $n=10^6{\rm cm}^{-1}$. 
The widths of the electron gas wire is $a=b=200\AA$.
}
\label{f2}
\end{figure}

\begin{figure}
\caption{ Intensity of phonon scattering as functions
of phonon frequency $\omega$ at both high temperature (solid-line) 
and low temperature (dotted-line), 
where $I(\omega)$ is defined in the text.
The dotted-line represents the low temperature result multiplied by 
a factor of $10^7$.
The density of the electron gas is $n=10^6{\rm cm}^{-1}$.
The widths of the electron gas wire is $a=b=200\AA$.
}
\label{f3}
\end{figure}

\begin{figure}
\caption{Resistivities of a quantum wire
from bare phonon scattering (dotted-line) and from
the renormalized phonon scattering (solid line) 
as functions of electron density at low temperature. 
The widths of the electron gas wire is $a=b=200\AA$.
The temperature is $k_BT=0.02\hbar\omega_{2k_F}$.
}
\label{f4}
\end{figure}


\begin{references}
\bibitem{sak1}
H. Sakaki, Jpn. J. Appl. Phys. {\bf 19}, L735 (1980).

\bibitem{1dr}
H. Bruus, K. Flensberg, and H. Smith,
Phys. Rev. B. {\bf 48}, 11144 (1993);
C.W.J. Beenakker and H. van Houten, Solid State Phys. {\bf 44} 1 (1991);
T. Martin and S. Feng, Phys. Rev. Lett. {\bf 64}, 1971 (1990);
D.L. Maslov, Y.B. Levinson, and S. M Badalian, Phys. Rev. B. {\bf 46},
7002 (1992);
S. Komiyama, {\it et. al.},    
Phys. Rev. B. {\bf 45}, 11085 (1992);
H. Akera and T Ando, Phys. Rev. B. {\bf 41}, 11967 (1990);
P. Vasilopoulos, {\it et. al.},  Phys. Rev. B. {\bf 40}, 1810 (1989);
N. Mori, H. Momose, and C. Hamaguchi, Phys. Rev. B. {\bf 45},
4536 (1992);  H. Bruus and K. Flensberg, J. Phys. Condens. Matter {\bf 4},
9131 (1992);
H.C. Tso and P. Vasilopoulos, Phys. Rev. B. {\bf 44}, 12952 (1991);
I.I. Boiko, P. Vasilopoulos, and V.I. Sheka,
Phys. Rev. B. {\bf 46}, 7794 (1992);
B.Y-K. Hu and S. Das Sarma, Appl. Phys. Lett. {\bf 61}, 1208 (1992);
V.L. Gurevich, V.B. Pevzner, and G. Iafrate, 
Phy. Rev. Lett. {\bf 75}, 1352 (1995).

\bibitem{pri}
P.J. Price, Solid State Comm. {\bf 51}, 607 (1984);
Ann. Phys. (NY) {\bf 133} 217(1981); Phys. Rev. B. {\bf 32}, 2643 (1985).

\bibitem{gp}
D. G\"otze and P. W\"olfle, Phys. Rev. B. {\bf 6}, 1226 (1972).

\bibitem{for}
D. Forster, {\it Hydrodynamic Fluctuations, Broken Symmetry, and
Correlation Functions} (Addison-Wesley, 1990).

\bibitem{find}
Y. Shiwa and A. Isihara, J. Phys. C {\bf 16}, 4853 (1983);
I.C. da Cunha Lima and S.C. Ying, J. Phys. C. {\bf 18}, 2887 (1983).
 
\bibitem{dc}
R.C. Dynes and J.P. Carbotte, Phys. Rev. {\bf 175}, 913 (1968).

\bibitem{lz}
L. Zheng and A.H. MacDonald, Phys. Rev. B. {\bf 48}, 8203 (1993).

\bibitem{hol}
T. Holstein, Ann. Phys. (NY)  {\bf 29}, 410 (1964).

\bibitem{1d1p}
M.A. Stroscio and K.W. Kim, Phys. Rev. B. {\bf 48}, 1936 (1993);
M.A. Stroscio, Phys. Rev. B. {\bf 40}, 6428 (1989);
M.A. Stroscio, {\it et al.}, Superlatt. Microstruct. 
{\bf 10}, 55(1991);
S. Das Sarma and V.B. Campos, Phys. Rev. B. {\bf 49}, 1867 (1994).

\bibitem{hus}
P.F. Maldague, Surf. Sci. {\bf 73}, 296 (1978);
B.Y-K. Hu and S. Das Sarma, Phys. Rev. B. {\bf 48}, 5469 (1993).

\end{references}
\end{document}